\documentclass[12pt,preprint]{aastex}

\shorttitle{Prediction of solar cycles }
\shortauthors{Hiremath}

\begin{document}

\title{Prediction of future fifteen solar cycles} 

\author{K. M. Hiremath}
\affil{Indian Institute of astrophysics, Bangalore-560034, India}
\email{hiremath@iiap.res.in}
\begin{abstract}
In the previous study (Hiremath 2006a), the solar cycle is modeled
as a forced and damped harmonic oscillator and from all the 22 cycles
(1755-1996), long-term amplitudes, frequencies, phases and decay
factor are obtained. Using these physical parameters 
of the previous 22 solar cycles and by an {\em autoregressive model},
 we predict the amplitude and period of the future fifteen solar cycles.
Predicted amplitude of the present solar cycle (23) matches very well with the observations.
The period of the present cycle is found to be 11.73 years.
With these encouraging results, we also predict the profiles of future 15 solar
cycles. Important predictions are : (i) the period and amplitude of the
cycle 24 are 9.34 years and 110 ($\pm 11$), (ii) the period and amplitude of
the cycle 25 are 12.49 years and 110 ($\pm$ 11),
 (iii) during the cycles 26 (2030-2042 AD), 27 (2042-2054 AD), 
34 (2118-2127 AD), 37 (2152-2163 AD) and 38 (2163-2176 AD), 
the sun might experience a very high sunspot activity,
(iv) the sun might also experience a very low ( around 60) sunspot activity
 during cycle 31 (2089-2100 AD) and, (v) length of the solar cycles
vary from 8.65 yrs for the cycle 33 to maximum of 13.07 yrs
for the cycle 35.

\end{abstract}


\section{Introduction}
 Owing to proximity, the sun influences the earth's climate
and environment. Overwhelming evidence is building up that
the solar cycle and related activity phenomena are correlated
with the earth's global climate and temperature, the sea surface
temperatures of the three (Atlantic, Pacific and Indian) main
ocean basins, the earth's albedo, the galactic cosmic ray flux
that in turn is  correlated with
the earth's cloud cover and, Indian monsoon rainfall (Hiremath and
Mandi 2004 and references there in; Georgieva {\em et. al.} 2005;
 Hiremath 2006b). The transient parts of the solar activity
such as the flares and the coronal mass ejections that are
directed towards the earth create
havoc in the earth's atmosphere by disrupting the 
global communication, reducing life time of the earth bound satellites
and, keep in dark places of the earth that are at higher latitudes by
breaking the electric power grids. 
Owing to sun's immense influence of space weather effects
on the earth's environment and climate, it is necessary
to predict and know in advance different physical
parameters such as amplitude and period of the future
solar cycles.

There are many predictions in the literature (Ohl 1966;
Feynman 1982; Feynman and Gu 1986; Kane 1999; Hathaway, Wilson and
Reichmann 1999; Badalyan, Obrido and Sykora 2001; Duhau 2003
Sello 2003; Maris, Poepscu and Besliu 2003; Euler and Smith 2004;
Maris, Poepscu and Besliu 2004; Kaftan 2004; 
Echer {\em et. al.} 2004; Gholipour 
{\em et. al.,} 2005; Schatten 2005; Li, Gao and Su 2005; 
Svaalgaard, Cliver and Kamide 2005; Chopra and Dabas 2006; Dikpati, Toma and Gilman 2006;
Du 2006; Hathaway and Wilson 2006; Clilverd {\em et. al.,} 2006; Tritakis
and Vasilis 2006; Lantos 2006; Lundstedt 2006; Wang and Sheeley 2006; 
Choudhuri, Chatterjee and Jiang 2007; Javaraiah 2007)
 on the previous and future 24th solar cycles and beyond. Most of these
studies mainly concentrate on prediction of the 
amplitude (maximum sunspot number during a cycle). However, prediction of period
(length) of a solar cycle is also very important parameter
and the present study fills that gap. 

Recently we modeled the solar activity cycle
as a forced and damped harmonic oscillator that consists of both
the sinusoidal and transient parts (eqn 1 of Hiremath 2006a).  
From the 22 cycles (1755-1996) sunspot data, the physical parameters 
(amplitudes, frequencies, phases and decay factors) of such a
harmonic oscillator are determined. The constancy of the amplitudes 
and the frequencies of the sinusoidal part
and a very small decay factor from the transient part suggests that
the solar activity cycle mainly consists of persistent oscillatory part
that might be compatible with long-period ($\sim  22 \,yrs$) Alfven oscillations. 
In the present study, with an {\em autoregressive} model 
and by using the physical parameters of 22 cycles, 
we predict the amplitudes and periods of future 16 solar cycles.
Thus prediction from this study can be considered
as a {\em physical and precursor} method.

A Pth order {\em autoregressive} model relates a forecasted value $x_{t}$ of the time 
series $X = [x_{0}, x_{1}, x_{2}, ... , x_{t-1}]$, as a linear combination of $P$ past values
 $x_{t} = \phi_{1} x_{t-1} + \phi_{2} x_{t-2} + ...... + \phi_{p}x_{t-p} + W_{t}$ ,
where the coefficients $\phi_{1}, \phi_{2}, ... , \phi_{p}$  are calculated such that they 
minimize the uncorrelated random error terms, $W_{t}$. The routine is
available in IDL software. Important condition for using
an {\em autoregressive model} is that the series must
be {\em stationary} such that it's mean and standard
deviation do not vary much with time. Hence, one can not
apply {\em autoregressive model} directly  to the observed sunspot series as
it consists of near sinusoidal trends whose amplitudes and the standard deviations 
entirely different for different solar cycles.
On the other hand, the derived physical parameters of
the forced and damped harmonic oscillator (Hiremath 2006a)
for all the 22 solar cycles
are stationary and, hence in the following,
we  use an {\em autoregressive} model to
predict the future 15 solar cycles. 

The solution of the forced and damped harmonic oscillator
(see the equation 1 of Hirmath 2006a)  of the 
solar cycle consists of two parts :
(i) the {\em sinusoidal} part that determines the amplitude
and period of the solar cycle and, (ii) the {\em transient}
part that dictates decay of the solar cycle
from the maximum year and also determines bimodal structure
of the sunspot cycle around the maximum years
for some cycles. In the present study, we use physical parameters
of the sinusoidal part only to predict amplitude and period of future cycles.  

\section{Results and conclusion}

Using past 22 cycles' physical parameters, we construct
the next (23rd)  solar cycle and presented in Fig 1. Except
decaying part of the solar cycle, one can notice that the predicted curve
exactly matches with the observed curve. With this encouragement and from an
{\em autoregressive} model, the physical parameters of future 16 solar cycles
are computed and reconstructed solar cycles are presented in
Fig 2. For the coming cycles 24-38, the results are summarized
in Table 1. In Table 1, the first column represents the cycle
number, the second column represents the year from minimum-minimum,
the third column represents the period (length) of the solar
cycle and, the last column represents the maximum
sunspot number during a cycle. It is interesting to note that the amplitude of the cycle 24
is low compared to the amplitude of cycle 23 and is almost 
similar to average value computed from all
of the predicted models ($http://members.chello.be/j.janssens/SC24.html$).    
Other interesting predictions are : (i) during the cycles 26, 27, 34. 37
and 38, the sun will experiences a very high solar activity, (ii) during
cycle 31 (2087-2099 AD) the sun will experiences a very low sunspot activity
and, (iii) length of the solar cycles vary from 8.65 yrs for the cycle 33 to
maximum of 13.07 yrs for cycle 25.  

To conclude, the solar cycle is modeled as a forced and damped harmonic
oscillator. From the previous 22 cycles sunspot data, the physical
parameters such as the amplitudes, the frequencies and phases of
such a harmonic oscillator are determined. The sinusoidal
part of the forced and damped harmonic oscillator of
previous solar cycles is considered for the prediction
of future 16 cycles. With an {\em autoregressive}
model and using previous 22 cycles parameters, coming 16 solar  cycles are 
reconstructed from the predicted parameters. Important results of this
prediction are : the amplitude of coming solar cycle 24 will be smaller
than the present cycle 23 and around 2087-2099 AD, the sun will experiences
a very low sunspot activity.  

\acknowledgments

The author is thankful to Dr. Luc Dame and Dr. Javaraiah for the useful discussions.

\clearpage

\begin{figure}[h]
{{  \noindent\includegraphics[width=15pc,height=15pc]{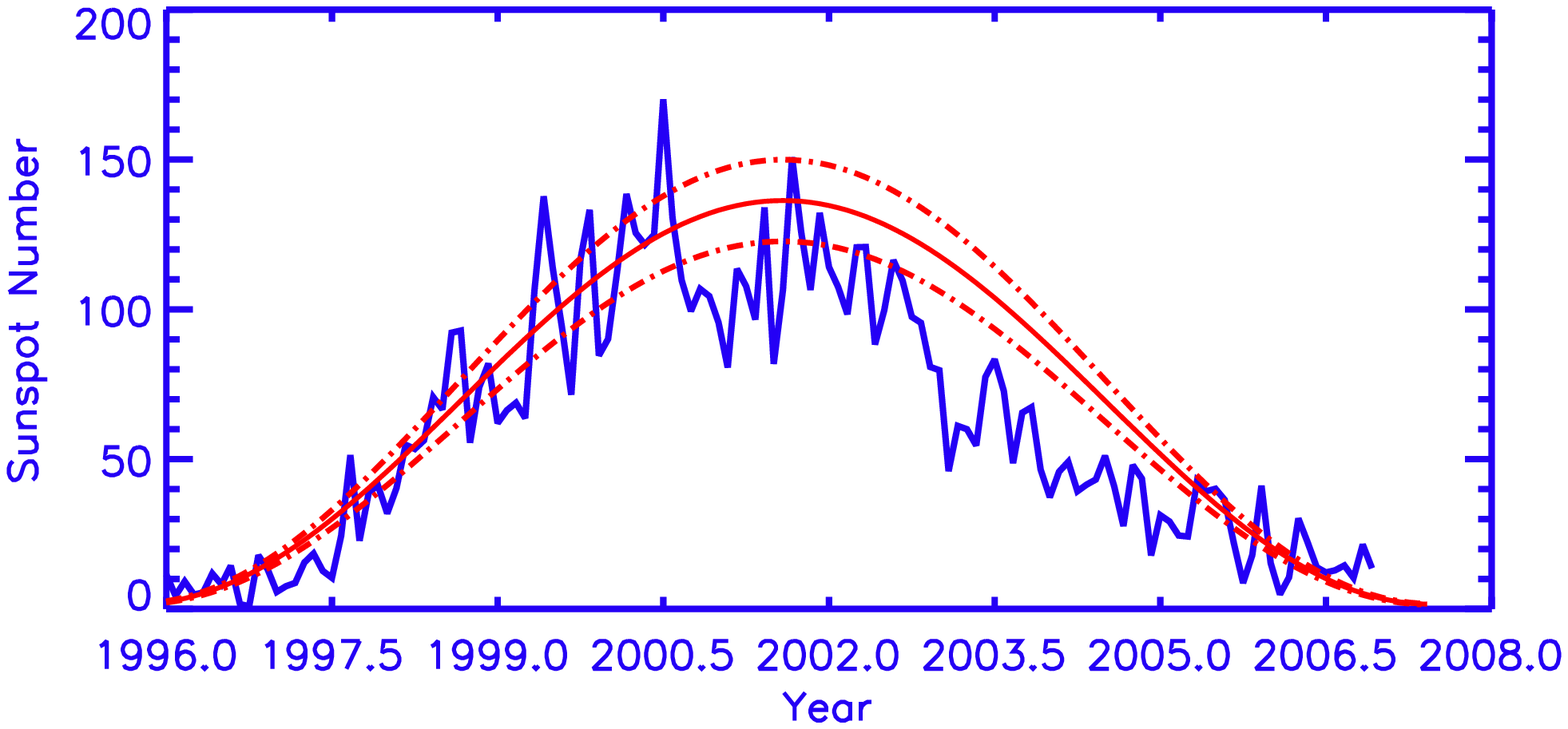}
  \noindent\includegraphics[width=25pc,height=15pc]{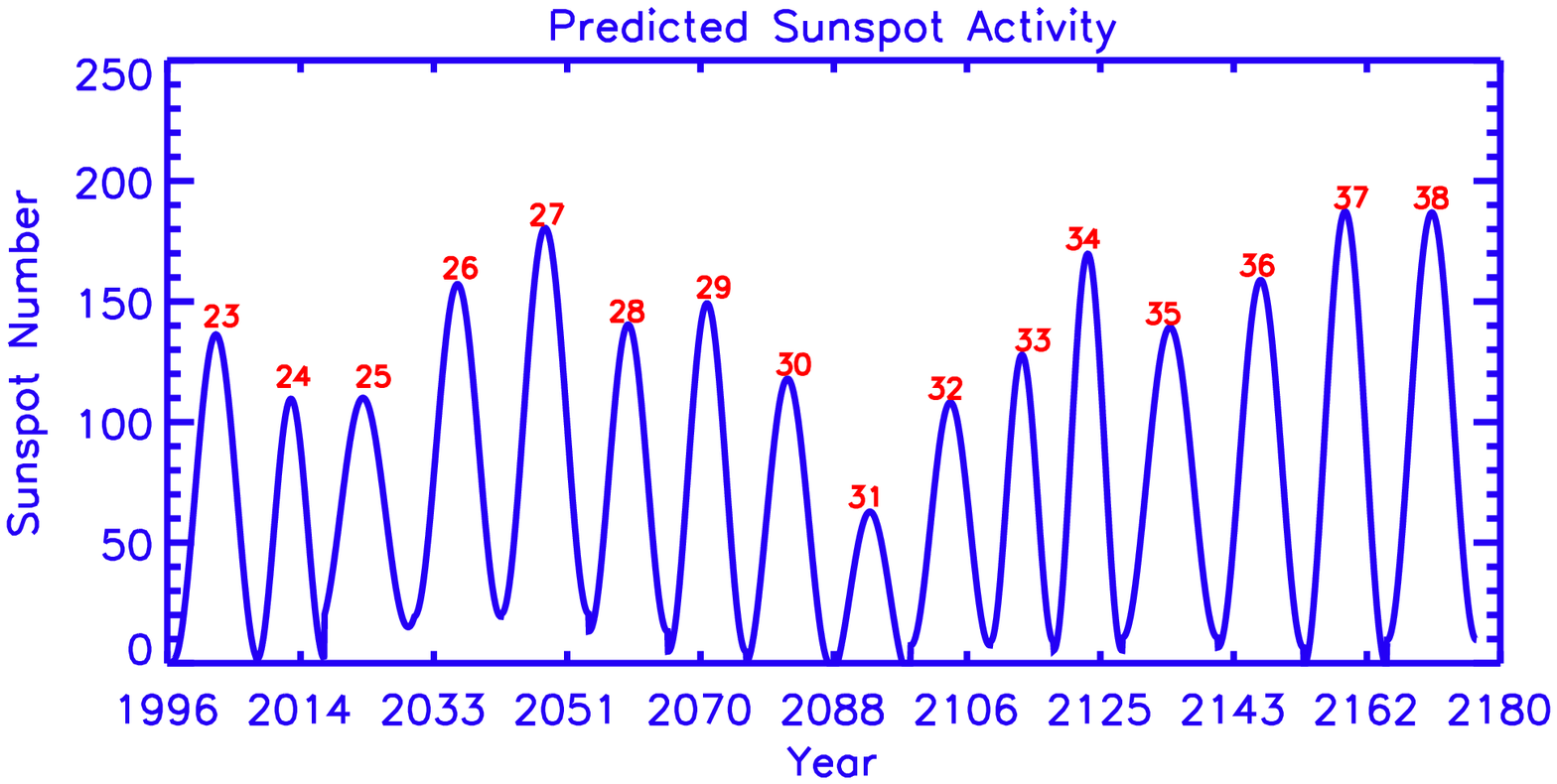}
}}
\caption{Prediction of the future solar cycles. (a) The left plot 
illustrates the predicted (red continuous) curve over plotted on the 
observed (blue curve) sunspot cycle 23. The dashed red curves represent
uncertainty in the Prediction. (b) The right plot 
illustrates predicted future 15 solar cycles. The red numbers over different
solar cycle maximum are the cycle numbers.}
\end{figure}
\begin{deluxetable}{rrrrrrrr}
\tablecolumns{5}
\tablewidth{0pc}
\tablecaption{ Predicted sunspot cycles}
\tablehead{
\colhead{Cycle} & \colhead{Year}   & \colhead{Period}    & \colhead{Maximum}\\
\colhead{Number} & \colhead{Min-Min}   & \colhead{(Years)}    & \colhead{Number}
}
\startdata
23\phn & 1996.00-2007.73\phn & 11.73 & 136$\pm$14 \\
24\phn & 2007.73-2017.07\phn & 9.34 & 110$\pm$11 \\
25\phn & 2017.07-2029.56\phn & 12.49 & 110$\pm$11 \\
26\phn & 2029.56-2041.50\phn & 11.94 & 157$\pm$16 \\
27\phn & 2041.50-2053.51\phn & 12.00 & 180$\pm$18 \\
28\phn & 2053.51-2064.30\phn & 10.80 & 140$\pm$14 \\
29\phn & 2064.30-2075.01\phn & 10.71 & 149$\pm$15 \\
30\phn & 2075.01-2086.79\phn & 11.78 & 118$\pm$12 \\
31\phn & 2086.79-2097.95\phn & 11.16 & 63$\pm$6 \\
32\phn & 2097.95-2108.84\phn & 10.89 & 108$\pm$11 \\
33\phn & 2108.84-2117.49\phn & 8.65 & 128$\pm$13 \\
34\phn & 2117.49-2126.92\phn & 9.43 & 170$\pm$17 \\
35\phn & 2126.92-2139.99\phn & 13.07 & 139$\pm$14 \\
36\phn & 2139.99-2151.74\phn & 11.75 & 159$\pm$16 \\
37\phn & 2151.74-2163.19\phn & 11.45 & 187$\pm$19 \\
38\phn & 2163.19-2175.48\phn & 12.29 & 187$\pm$19 \\
\enddata
\end{deluxetable}

\end{document}